\documentclass[pre,twocolumn,showpacs,superscriptaddress,showkeys]{revtex4}
\usepackage{dcolumn}
\usepackage{bm}
\usepackage{amssymb}
\usepackage{graphicx,epsfig}
\begin{document}

\title{Pair production by a circularly polarized electromagnetic wave in plasma.}
\author{S. S. Bulanov}
\affiliation{Institute of Theoretical and Experimental Physics, B.
Cheremyshkinskaya str. 25, Moscow 117218, Russia}
\email{bulanov@heron.itep.ru}

\begin{abstract}
We present the results of the calculation of the electron-positron
pair creation probability by the circularly polarized
electromagnetic wave during its propagation in underdense
collisionless plasmas. The dependence of the probability on the
frequency and the amplitude of the pulse is studied in detail.
\end{abstract}



\maketitle

\section{Introduction}

Quantum Electrodynamics (QED) predicts the possibility of
electron-positron pair production in a vacuum by strong electric
field \cite{Schwinger}. This nonlinear effect has attracted great
attention due to the fact that it lays outside the scope of
perturbation theory and sheds a light on the nonlinear quantum
electrodynamics properties of the vacuum \cite{vacuum}. As was
shown in Refs. \cite{Schwinger, el-pos} a planar electromagnetic
wave of arbitrary intensity and spectral composition does not
produce the electron-positron
pairs in a vacuum because it has both the electromagnetic field invariants $%
J_{1}=({\bf B}^{2}-{\bf E}^{2})/2$ and $J_{2}=({\bf E\cdot }{\bf %
B})/2$ equal to zero \cite{LandauLifshits-TF}. Due to this reason the pair
creation was first considered for a static electric field, which evidently
has non-vanishing invariant $J_{1}$, then its theoretical description was
extended on the time-varying electric-type electromagnetic field \cite{B-I}.
In this case the electromagnetic field invariants $J_{1}$ and $J_{2}$ obey
conditions
\begin{eqnarray}
J_{1} &=&\frac{1}{4}F_{\mu \nu }F^{\mu \nu }=\frac{1}{2}({\bf B}^{2}-%
{\bf E}^{2})<0,  \nonumber \\
J_{2} &=&\frac{1}{4}F_{\mu \nu }{\tilde{F}}^{\mu \nu }=\frac{1}{2}({\bf %
E\cdot }{\bf B})=0,
\end{eqnarray}
where $F_{\mu \nu }$ is the electromagnetic field tensor and
${\tilde{F}}^{\mu \nu
}=\varepsilon^{\mu\nu\rho\sigma}F_{\rho\sigma}$,
$\varepsilon^{\mu\nu\rho\sigma}$ is the totally antisymmetric
tensor. In particular, the pair creation in a vacuum in a
spatially homogeneous electric field
\begin{equation}
{\bf E}(t)=F\phi (t){\bf e}_{z},~~~~{\bf B}(t)=0,  \label{Et}
\end{equation}
where $\phi (t)=\cos t$ and ${\bf e}_{z}$ is a unit vector in the
z-direction, was considered in Refs.
\cite{B-I,3,4,PopovMarinov,NN,7,8,9,Ringwald}.

The process (called the Schwinger pair production mechanism) is of
fundamental importance in QED and in quantum field theory in
general. However, previous estimates \cite{4,25,26} showed, that
it hardly could be observed in the experiments using by that time
available optical lasers whose power was far from the limit able
to provide an intensity of the order of $10^{29}$W/cm$^{2}$, which
corresponds to the Schwinger critical electric field $1.32\times
10^{16}$~V/cm. As closer the electric field approaches the Schwinger
limit as the sub-barrier tunneling results in more efficient electron-positron
pair production. Therefore, the results of studies presented in
Refs. \cite{Schwinger,B-I,3,4,PopovMarinov,NN,7,8} for a long time
were generally believed to be of purely theoretical interest in
QED.

Recently, fortunately, the situation has drastically changed,
because the power of optical and infrared lasers rose by many
orders of magnitude \cite {27}. This opened several ways to
approach critical QED intensity. One way was demonstrated in the
SLAC experiments \cite{SLAC99}, where 46.6 GeV electron beam
interacted with counterpropagating laser pulse, which intensity
was 0.5$\cdot 10^{18}$W/cm$^{2}$. The incident radiation in its turn interacted with the similar
laser pulse and several electron-positron pairs were detected. Then, the projects
to build free-electron lasers at TESLA electron-positron collider
in DESY and the corresponding facilities at SLAC, in which
coherent laser beams with the photon energies of the order of
several keV are supposed to be produced, are being designed
\cite{29} (see also Refs. \cite{Ringwald,FEL,Tajima}). Recently, an approach
to the light intensification towards the Schwinger limit by using nonlinear
interaction between the electromagnetic and the langmuir wave in plasmas
has been formulated in Ref. \cite{sbul}. In this scheme the QED critical field
can be achieved with present day laser systems. Hence the
theory of the Schwinger effect must be considered in more detail
in view of the new experimental possibilities.

However, here arises a question, whether the model of pair
creation by the time-varying electrical field (\ref{Et}) in a
vacuum resembles well enough the experimental situation. Such a
field can be generated in the antinodes of the standing light
wave, which is realized in a superposition of two
counter-propagating laser beams. In this case, the region where
electric field is much larger than the magnetic field, is
relatively small and the results obtained in Refs.
\cite{B-I,3,4,PopovMarinov,NN,7,8,9,Ringwald,Avetissian} describe just a
small portion of the pairs that could be created by the
electromagnetic wave. In the rest part, the pair creation will be
affected by the presence of the time-varying magnetic field. This
in its turn should lead to very complicated formulas for pair
production probability.

To elucidate the role of the magnetic field component on the
electron positron creation we consider below the planar circularly
polarized electromagnetic wave propagating in the underdense
collisionless plasma. Even in a very low density plasma the first
electromagnetic invariant $J_{1}$ calculated for planar
electromagnetic wave is not equal to zero. That is why in a plasma
the electron-positron pairs can be created by the planar wave. In this case the
usage of the Lorentz transformation into the reference
frame moving with the wave group velocity leads to the electromagnetic  wave
field transformation to the purely electric field,
that rotates with constant frequency, and with no magnetic field.
Although this anzatz reduces the problem under consideration to
the situation when the pairs are created by the time varying
electric field, actually, the wave magnetic field effects are
incorporated rigorously into our model. We notice that similar approach has been earlier used
in Ref. \cite{Av}. In the present paper we use the method
of "imaginary time", simple and powerful mathematical technique for analytical
continuation of the classical equations of motion postulated to apply inside the
barrier, through the barrier where classical motion is not allowed to join up with the
equations of motion for particles  that have escaped through the barrier to the outside where
classical motion is again allowed (see Refs. \cite{PTP,Popov}). With the help of ''the
imaginary time technique'' we shall study in details
the dependence of the pair creation probability on the frequency
and the amplitude of the wave.

In addition to the pairs created by the electromagnetic wave via
the Schwinger mechanism there is electron-positron pair creation
due to the trident process \cite{trident} and bremsstrahlung
photons. We shall show that the electron-positron pair produced
via the Schwinger mechanism can be distinguished from the one
appeared due to other mechanisms by the momentum filter, i. e. the
momentum of the "Schwinger" pair in the laboratory frame is
larger.

The paper is organized as follows. In section 2 we discuss the properties of
the relativistically strong electromagnetic wave in plasma. The probability
of pair creation by such a wave is calculated in section 3. In section 4 the
discussions of main results and conclusions are presented.

\section{Relativistically Strong Electromagnetic Wave in Plasma}

In this Section we recover well known properties of a
relativistically strong electromagnetic wave in an underdense
collisionless plasma needed for further formulation of the time
varying electric field configuration. A discussion of the wave
behavior is based on the results obtained by Akhiezer and Polovin
in Ref. \cite{AP}. We consider the circularly polarized
electromagnetic wave, which is described by the transverse
component of the vector potential:

\begin{equation}
{\bf A}_{\bot }=A_{0}\left[ {\bf e}_{y}\sin (\omega t-kx)-{\bf e}%
_{z}g\cos (\omega t-kx)\right] .  \label{A}
\end{equation}
The wave propagates in the direction of the $x$ axis, its phase velocity
equals $\omega /k$, $g$ represents right-handed ($g=1$) or left-handed ($g=-1
$) circular polarization, and $A_{0}=cE_{0}/\omega $ with $E_{0}$ being the
amplitude of the electric field. The dependence of the wave frequency $%
\omega $ on the $x$-component of the wave-vector $k$ is given by the
dispersion equation
\begin{equation}
\omega ^{2}=k^{2}c^{2}+\sum\limits_{\alpha }\frac{\omega _{p\alpha }^{2}}{%
\left[ 1+(Z_{\alpha }eA_{0}/m_{\alpha }c^{2})^{2}\right] ^{1/2}},  \label{DE}
\end{equation}
where $\alpha =e,p,i,...$ denotes species in the plasma with the electric
charge $Z_{\alpha }e$ , mass $m_{\alpha }$, and density $n_{\alpha }$, $%
\omega _{p\alpha }=\left( 4\pi n_{\alpha }Z_{\alpha }^{2}e^{2}/m_{\alpha
}\right) ^{1/2}$, and $c$ is a speed of light in a vacuum. By virtue of the
plasma electric neutrality condition we have $\sum\limits_{\alpha }Z_{\alpha
}n_{\alpha }=0$. Introducing notation for the frequency $\omega $%
\begin{equation}
\Omega ^{2}=\sum\limits_{\alpha }\frac{\omega _{p\alpha }^{2}}{\left[
1+(Z_{\alpha }eA_{0}/m_{\alpha }c^{2})^{2}\right] ^{1/2}}, \label{Omega}
\end{equation}
we rewrite the dispersion equation in the form $\omega =(k^{2}c^{2}+\Omega
^{2})^{1/2}$.

The electric and the magnetic field are equal to
\begin{equation}
{\bf E}=\frac{1}{c}\partial _{t}{\bf A}_{\bot }=\frac{\omega}{c} A_{0}[{\bf e%
}_{y}\cos (\omega t-kx)+{\bf e}_{z}\sin (\omega t-kx)]
\end{equation}
and
\begin{equation}
{\bf B}=\nabla \times {\bf A=}kA_{0}[{\bf e}_{y}\sin (\omega t-kx)+%
{\bf e}_{z}\cos (\omega t-kx)],
\end{equation}
respectively.

We see that in a plasma the first invariant of the electromagnetic field $%
J_{1}$ is not equal to zero:
\begin{equation}
J_{1}=\frac{1}{2}(B^{2}-E^{2})=-\frac{\Omega ^{2}}{2c^{2}}A_{0}^{2}\equiv
-\left( \frac{\Omega }{\omega }\right) ^{2}E_{0}^{2}.
\end{equation}
It vanishes when the plasma density tends to zero, i. e. in a vacuum (or in
the limit $A_{0}\rightarrow \infty $). It also vanishes when the charges of plasma
species go to zero, because $\Omega\sim \omega_{p\alpha}$, which in its turn
$ \omega_{p\alpha}\sim Z_\alpha$.

The nonlinear electromagnetic wave in plasmas is also
characterized by the dependence of its phase and group velocity on
the plasma parameters and on the wave amplitude. From equation
(\ref{DE}) we find that the phase velocity
$v_{ph}=\omega /k$ and the group velocity $v_{g}=\partial \omega /\partial k$%
, are related to each other via expression $v_{ph}v_{g}=c^{2}$.

Now we perform the Lorentz transformation to the reference frame moving with
the group velocity along the direction of the wave propagation. The wave
frequency and the $x-$component of the wave vector change according to
\begin{equation}
\omega ^{\prime }=\frac{\omega -kv_{g}}{(1-v_{g}^{2}/c^{2})^{1/2}}\qquad
\mathrm{and}\qquad k^{\prime }=\frac{k-\omega v_{g}/c^{2}}{%
(1-v_{g}^{2}/c^{2})^{1/2}}.
\end{equation}
Using relationship $v_{g}=c^{2}/v_{ph}=kc^{2}/\omega $, we obtain
that in the moving reference frame the wave frequency equals
$\omega ^{\prime }=\Omega $ and its wave number vanishes:
$k^{\prime }=0$. Using the Lorentz invariance of the transverse
component of the vector potential we find that in the moving
reference frame the magnetic field in the circularly polarized
electromagnetic wave vanishes and the electric field, rotating
with the frequency $\Omega $, is given by expression:
\begin{equation}
{\bf E}=\frac{\Omega}{c} A_{0}\left( {\bf e}_{y}\cos \Omega t^{\prime }+g%
{\bf e}_{z}\sin \Omega t^{\prime }\right) .  \label{Etr}
\end{equation}
Further we shall use a notation $E=\left(\Omega
A_{0}/c\right)\equiv (\Omega /\omega )E_{0}$. See also Ref.
\cite{Av}, where the model case of linearly polarized wave has
been considered. The finite amplitude linearly polarized wave
propagating in plasmas has both transversal and longitudinal
components, oscillating with different frequencies. That is why
the problem of the pair production by this wave is technically
more complicated, than the case of circularly polarized wave. It
will be considered in our forthcoming publications.

\section{Pair Production by Circularly Polarized Wave}

We consider the problem of pair production in the circularly polarized
electric field, using the approach formulated in Ref. \cite{PopovMarinov}.
The wave function of the electron in the electromagnetic field has the form
\begin{equation}
\psi _{f}({\bf p},t)=\int d^{3}r~d^{3}r_{0}e^{-ipr}G(r,t;r_{0},0)\psi
_{i}(r_{0},0).  \label{wfunc}
\end{equation}
The Green function, according to Feynman, in the coordinate
representation has the following asymptotic behavior
\begin{equation}
G(r,t;r_{0},0)\sim (2\pi it)^{3/2}\exp [iS(r,t;r_{0},0)],
\end{equation}
where $S$ is the action for the electron.
For the motion in the in the homogeneous field $\psi _{i}\sim \exp (i{\bf %
p}_{0}{\bf r}_{0})$. Integration of Eq. (\ref{wfunc}) with the use of saddle point method
yields the following equations for saddle point coordinates (see Ref. \cite{PopovMarinov}):
\begin{equation}
\frac{dS}{dr}-p=0,~~~\frac{dS}{dr_{0}}+p_{0}=0.
\end{equation}
As it is known in the classical mechanics these equations
determine the extremal trajectory, which satisfies the Lagrange
equations. Therefore with the accuracy up to the preexponential
factor we get
\begin{equation}
\psi _{f}\sim \exp \left(i\hat{S}\right)
\end{equation}
with
\begin{equation}
\hat{S}=\int\limits_{0}^{t}\mathcal{L}dt-{\bf p\cdot r}+{\bf p}_{0}{\bf %
\cdot r}_{0}, \label{W}
\end{equation}
where $\hat{S}$ is the reduced action \cite{PopovMarinov}, ${\bf
r}$ is the coordinate of the particle, and ${\bf r}_{0}$ is the
initial coordinate and the Lagrangian (see Ref. \cite
{LandauLifshits-TF}) is given by expression
\begin{equation}
\mathcal{L}=-m_{e}c^{2}\left( 1-|{\bf v|}^{2}/c^{2}\right) ^{1/2}+e%
{\bf A\cdot v}-e\varphi ,
\end{equation}
which depends on the vector ${\bf A}$ and the scalar $\varphi $
potential of the electromagnetic field and on the charged particle velocity $%
{\bf v}=c{\bf p}/(m_{e}^{2}c^{2}+|{\bf p|}^{2})^{1/2}$. Here $%
{\bf |p|}^{2}=p_{x}^{2}+p_{y}^{2}+p_{z}^{2}$. In particular case, when the
charged particle interacts with the electromagnetic wave, the integrand in
the action functional (\ref{W}) can be expressed via the particle energy $\mathcal{E}%
=\left( m_{e}^{2}c^{4}+|{\bf p|}^{2}c^{2}\right) ^{1/2}$ as
\begin{equation}
\hat{S}=-\int\limits_{0}^{t}\mathcal{E}dt.
\end{equation}
An expression for the action remains unchanged in a plasma, if we assume
that ${\bf A}$ and $\varphi $ are the vector and scalar potentials of net
electromagnetic field acting on the electron or positron. The representation
of the electromagnetic wave field in the form (\ref{Etr}) with the frequency $\Omega$ given
by Eq. (\ref{Omega}) implies the use of the macroscopic description of the wave. The analysis
of the microscopic structure of the electron-positron pair production in plasmas is
beyond the scope of the present publication. However, since similar technique of the Lorentz
transformation into the reference frame, moving with the wave group velocity,
can be used to simplify the analysis of electron-positron pair
production inside the hollow waveguide and in the focus region \cite{focus}, we think  that
the main results obtained below in the longwave approximation ($\left(2\pi c/\omega\right)^3 n \gg 1$)
will not change drastically.

The probability of pair creation depends on the imaginary part of the action as
\begin{equation}
dw_{p}\sim \exp
\left\{-2~\mbox{Im}\left[\hat{S}(p)\right]\right\}d^{3}p.
\end{equation}
To calculate the pair creation probability we need to find the extremal
trajectory for the sub-barrier motion of the particle, which minimizes the
action. The action functional variation is
\begin{eqnarray}
\delta \hat{S} &=&-\int\limits_{0}^{t}dt\left( \frac{\partial \mathcal{E}}{%
\partial p_{i}}\delta p_{i}+\frac{1}{2}\frac{\partial ^{2}\mathcal{E}}{%
\partial p_{i}\partial p_{j}}\delta p_{i}\delta p_{j}+...\right)   \nonumber
\\
&=&a_{i}\delta p_{i}+\frac{1}{2}b_{ij}\delta p_{i}\delta p_{j}
\end{eqnarray}
with the functions
\begin{equation}
a_{i}=-\int\limits_{0}^{t}dt\frac{p_{i}}{\mathcal{E}}=-\rho
_{i};~~~b_{ij}=-\int\limits_{0}^{t}\frac{dt}{\mathcal{E}}\left( \delta _{ij}-%
\frac{p_{i}p_{j}}{\mathcal{E}}\right).
\end{equation}
Here ${\bf \rho }={\bf r}(t)-{\bf r}_0$ is the particle
displacement during its sub-barrier motion. The requirement of
$\mbox{Im} [\hat{S}]$ being minimal leads to the condition
$\mbox{Im} [{\bf \rho}]=0$, which determines the extremal
trajectory. We introduce a notation ${\bf q}_{0}$ for
characteristic momentum, which corresponds to the extremal
trajectory. In the case, when the quasiclassical approximation is
valid, the momentum spectrum of emerging
from under the barrier particles has a sharp peak near ${\bf p}={\bf q}%
_{0}$ \cite{PopovMarinov}. However the discussion of the exact
form of the spectrum of the produced electrons and positrons is beyond the
scope of the present paper and it will be carried out elsewhere.
The sub-barrier motion is determined by the classical equations:
\begin{equation}
\dot{{\bf p}}=e{\bf E},
\end{equation}
\begin{equation}
\dot{{\bf r}}=c\frac{{\bf p}}{(m_{e}^{2}c^{2}+|{\bf p|}^{2})^{1/2}},
\end{equation}
We should add to these equations of sub-barrier motion the
requirement of trajectory to be the extremal one, i.e. $\mbox{Im} [{\bf \rho
}]=0$, then for the circularly polarized electric field of the
form
\begin{equation}
E_{x}=0,~~~E_{y}=E\cos \Omega t,~~~E_{z}=E\sin \Omega t,
\end{equation}
we find
\begin{equation}
p_{y}=q_{y}+P\sin \Omega t,~~~p_{z}=q_{z}-P\cos \Omega t,~~~p_{x}=q_{x},
\end{equation}
\begin{equation}
\mbox{Im} [{\bf \rho }]=\frac{2}{\Omega }\mbox{Re}\left[ \int\limits_{0}^{\tau _{0}}\frac{%
{\bf p}}{(m_{e}^{2}c^{2}+|{\bf p|}^{2})^{1/2}}d\tau\right] =0,
\end{equation}
where $\tau =-i\Omega t$ and $P=eE/\Omega $. From condition $\mbox{Im} [{\bf \rho }]%
=0$ we obtain $q_{y}=q_{x}=0$. As we see, when the particle
emerges from under the barrier ($t=0$) its momentum is
perpendicular to the instantaneous direction of the electric
field. Here we encounter the drastic difference of the two
dimensional case of the particle motion (it was first pointed out
in \cite{PopovMarinov}), which is realized in the circularly
polarized wave, from the one dimensional motion considered in
Refs. \cite{4,Ringwald,Popov}. In the former case the point of
particle emerging from under the barrier is the turning point of
the trajectory and therefore the particle momentum should be here
equal to zero. In the two dimensional case the component of
particle momentum along the instantaneous direction of the
electric field is zero. It is due to the fact that under the
barrier this component is purely imaginary. The perpendicular
component of the particle momentum
can be complex. We use a notation $q_{z}=Ps$ and rewrite the equations for $%
\tau _{0}$ and $s$ in the following form, which corresponds to the
equations of Ref. \cite{PopovMarinov} with ellipticity equal to
one:
\begin{equation}
\sinh ^{2}\tau _{0}-(s-\cosh \tau _{0})^{2}=\gamma ^{2},  \label{st0-a}
\end{equation}
\begin{equation}
\int\limits_{0}^{\tau _{0}}\frac{s-\cosh \tau }{\left[ \gamma ^{2}+(s-\cosh
\tau )^{2}-\sinh ^{2}\tau \right] ^{1/2}}d\tau =0,  \label{st0}
\end{equation}
where $\tau _{0}$ is the singular point of $W(t)$ with $W(t)$
being the action for the article moving along extremal trajectory.
$\tau _{0}$ has a meaning of sub-barrier motion time and $m_{e}
c/P=\gamma $. The parameter $\gamma$ was introduced in Refs.
\cite{B-I,3}, and plays a role of the adiabaticity parameter, as
is easily inferred from the function $g(\gamma)$, defined below,
which enters the principal exponential factor in the pair
production probability. Indeed, as long as $\gamma \ll 1$, i. e.
in the high-field, low-frequency limit the result obtained in the
framework of the imaginary method agrees with the nonperturbative
result of Ref. \cite{Schwinger} for a static, spatially uniform
field. Note that $\gamma$ characterizes the dynamics of particle
tunneling through a time varying barrier and is similar to the
well-known Keldysh parameter in the theory of the multiphoton
ionization of atoms and ions by laser radiation \cite{keldysh}.
Due to the fact that in the present paper we do not use Lorentz
factor, the notation $\gamma$ for the adiabaticity parameter
should not cause any ambiguity. The sub-barrier motion time $\tau
_{0}$ is determined by condition
$\mathcal{E}=(m_{e}^{2}c^{4}+|{\bf p|}^{2}c^{2})^{1/2}=0$, which
leads to equation (\ref{st0-a}). Equation (\ref{st0}) follows from
a condition $\mbox{Im} \left[\rho _{z}\right]=0$. The behaviour of
the functions $s$ and $\tau _{0}$ is presented in Figs. 1 a) and
b).

Using $\tau _{0}$ and $s$ from Eqs. (\ref{st0-a}) and (\ref{st0})
we can calculate the differential probability of the pair
creation. It reads (see Refs. \cite{prob1,prob2})
\begin{equation}
dw_{p}=e^{ \left\{ -\frac{\pi }{\varepsilon
}\left[g(\gamma)+c_x\frac{p_x^2}{m_e^2}+c_\perp\frac{(q-p_\perp)^2}{m_e^2}
\right]\right\}} d^3p,
\end{equation}
where $\varepsilon =E/E_{cr}$ is the normalized amplitude of the
electromagnetic wave in the moving reference frame, $p_\perp$ is
the momentum of the pair in the plane perpendicular to the
direction of wave propagation. $E_{cr}=2m_{e}^{2}c^{3}/e\hbar $ is
the Schwinger field, and the functions $g(\gamma )$, $c_x(\gamma
)$, $c_y(\gamma )$, and $c_z(\gamma )$ are given by expressions
\cite{PopovMarinov}:
\begin{equation}
g(\gamma )=\frac{4}{\pi \gamma }\int\limits_{0}^{\tau _{0}}
K^{1/2}d\tau ,
\end{equation}
\begin{equation}
c_x(\gamma )=g(\gamma)+\frac{\gamma}{2}\frac{d
g(\gamma)}{d\gamma},
\end{equation}
\begin{equation}
c_\perp(\gamma )=-\gamma\frac{d c_x(\gamma )}{d \gamma},
\end{equation}
where $K=1-\left[\sinh ^{2}\tau -(s-\cosh \tau )^{2}\right]/\gamma
^{2}$. According to Ref. \cite{Popov} the function $g(\gamma)$
demonstrates typical behavior of the pair creation in the linearly
polarized electric field. The function $g(\gamma)$ is presented in
Fig. 2 along with two typical examples from Ref. \cite{Popov}.

The total probability of pair creation can be represented as a sum of
probabilities of many-photon processes
\begin{equation}
W=\sum\limits_n w_n,
\end{equation}
where $w_n$ is $dw_p$ integrated over $d^3p$ with energy
conservation in the multiphoton process taken into account
(see Refs. \cite{3,4}). In the limit of small adiabaticity parameter ($\gamma\ll 1$), typical for  nowadays
lasers, \cite{Popov}, we get for the total probability of pair creation per unit volume per second
\begin{equation}
W\approx \frac{c}{4\pi^3 \lambda_e^4}\varepsilon^2\exp\left[-\frac{\pi}{%
\varepsilon}g(\gamma)\right].
\end{equation}
We should note, that unlike the case of linear polarization here
there is no additional factor $\sqrt{\varepsilon}$ (see
\cite{Popov} for details) compared to a constant field. It is due
to the fact, that in the limit $\omega\rightarrow 0$ the
circularly polarized field becomes constant one.

However the parameter $\gamma$ is not convenient for studying the
probability of pair creation, because the probability depends on
$\varepsilon$ and $\gamma$, which are not independent. These
parameters are related as
\begin{equation}
\gamma=\frac{m_e c}{P}=\frac{m_e c \omega}{e E}=\frac{2\pi
\lambda_e}{\lambda \varepsilon},
\end{equation}
where $\lambda_e=3.86\times 10^{-11}$ cm is the electron Compton length.
Therefore it is more convenient to study the
probability dependence on the electromagnetic wave wavelength, $\lambda$,
and on the parameter $\varepsilon$. In addition,
the electromagnetic wave is described in terms of amplitude
($\varepsilon$), wavelength ($\lambda$) and its shape, which
determines function $g(\gamma)$.

In order to understand at what wavelength the shape of the pulse
begins to play an important role, we present the dependence of
$g(\varepsilon)$ on parameter $\varepsilon$ in Fig. 3a for different values
of $\lambda$. It can be seen form this figure that for
$\lambda<10^{-7}$ cm the dependence of $g(\varepsilon)$ on
$\varepsilon$ should be taken into account. However all the
nowaday lasers are far from this limit (see Ref. \cite{Popov}).
This effect can manifest itself only in future X-ray lasers.
Moreover there is another problem with detecting this effect. One
should has rather small $\varepsilon$ (see Fig. 3a), at which the
pairs are produced at very small rate. As it can be seen from Fig.
3b, where the dependence of pair creation probability on
$\varepsilon$ for different values of $\lambda$ is shown, for
$\lambda=10^{-7}$ cm one pair is created at $\varepsilon\sim 0.08$
in the volume $(10^{-6}\mbox{cm})^3$ and in 500 fms. To elucidate
this point we present Figs. 3c and 3d, where the dependence of $g$
and $W$ on $\lambda$ for $\varepsilon=0.08$ are shown
respectively.

We should notice that the initial parameters of the
electromagnetic wave in vacuum enter the expression for pair
creation probability through parameters $\gamma $ and $\varepsilon
$:
\begin{equation}
\gamma =\frac{m_e c}{P}\frac{\omega }{\Omega }~~~~\mbox{and}~~~~\varepsilon =%
\frac{E}{E_{cr}}=\frac{E_{0}}{E_{cr}}\frac{\Omega }{\omega },
\end{equation}
where $E_{0}$ is the amplitude of the electric field in the
laboratory reference frame. So the electric field amplitude
effectively decreases in plasma and $\gamma $ increases.

In the reference frame moving with the wave group velocity using expression (%
\ref{Etr}) we find that the electron momentum is equal to
\begin{equation}
{\bf p}^{\prime }=P\left[ {\bf e}_{y}(s-\cos \Omega t^{\prime })-%
{\bf e}_{z}\sin \Omega t^{\prime }\right] .  \label{ptr}
\end{equation}
The electron energy is
\begin{equation}
\mathcal{E}^{\prime }=\left\{ m_{e}^{2}c^{4}+2P^{2}[1+\frac{s^{2}}{2}-s\cos
\Omega t^{\prime }]\right\} ^{1/2}.
\end{equation}
Performing the Lorentz transformation to the laboratory frame we obtain the
electron energy
\begin{equation}
\mathcal{E}=\frac{\omega }{\Omega }\left[ m_{e}^{2}c^{4}+2P^{2}\left(1+\frac{s^{2}%
}{2}-s\cos \chi \right)\right] ^{1/2}
\end{equation}
with $\chi =\omega t-kx$ and its longitudinal momentum
\begin{equation}
p_{x}=\left\{ \left[ \left( \frac{\omega }{\Omega }\right) ^{2}-1\right]
\left[ m_{e}^{2}c^{4}+2P^{2}\left(1+\frac{s^{2}}{2}-s\cos \chi\right )\right] \right\}
^{1/2}.
\end{equation}
The transverse momentum is given by
\begin{equation}
{\bf p}=\frac{P}{c}\{{\bf e}_{y}[s-\cos \chi]-{\bf e}_{z}\sin
\chi\}.
\end{equation}

In Fig. 4 we present averaged energy and longitudinal momentum versus $%
\gamma $. The averaging was performed on the phase of the pair creation,
i.e. by calculating integral
\begin{equation}
<\cdots >=\frac{\omega }{2\Omega \pi }\int\limits_{0}^{2\pi }(\cdots )d\chi .
\end{equation}
We see that in the limit of large $P$ and $\omega /\Omega $ the energy and
the longitudinal component of the particle are proportional to $%
m_{e}c^{2}(\omega P/\Omega )$ and $m_{e}c(\omega P/\Omega )$.

In addition to the pairs created by the electromagnetic wave via
the Schwinger mechanism there is electron-positron pair creation
due the trident process (see Ref. \cite{trident}). The electron
(positron) is created by the trident process in the laboratory
reference frame with the negligible small transverse momentum and
then it gains the transverse momentum equal to
\begin{equation}
{\bf p}^{\prime }=\frac{P}{c}\left[ {\bf e}_{y}(\cos \Omega t^{\prime
}-\cos \Omega t_{0}^{\prime })-{\bf e}_{z}(\sin \Omega t^{\prime }-\sin
\Omega t_{0}^{\prime })\right] .
\end{equation}
Contrary to the particles created via the Schwinger mechanism these
electrons (positrons) have in the moving reference frame non-zero
longitudinal momentum. It is equal to
\begin{equation}
p_{x}^{\prime }=-\frac{m_{e}v_{g}}{\left( 1-v_{g}^{2}/c^{2}\right) ^{1/2}}%
=-m_{e}c\left[ \left( \frac{\omega }{\Omega }\right) ^{2}-1\right] ^{1/2}.
\end{equation}
In Fig. 4b we present the dependence of the longitudinal momenta
in the laboratory frame of trident electrons (lower curve) and
''Schwinger'' electrons on parameter $\gamma $. It is obvious that
these two mechanisms of pair production can easily be
distinguished, because the longitudinal momentum of ''Schwinger''
electrons is larger. The same analysis can also be applied to the
electron-positron pairs produced by the bremsstrahlung photons
\cite{QED}, interacting with plasma particles.

We should note that one of the possible explanations of this
phenomenon is that "Schwinger" pairs are produced on vacuum, while
in all other mechanisms the pairs are produced on moving plasma
particles.

\section{Conclusions}

In the present paper we considered the problem of the
electron-positron pair creation by the circularly polarized laser
pulse in a plasma via the Schwinger mechanism. The representation
of the electromagnetic wave field in the form (\ref{Etr}) with the
frequency $\Omega$ given by Eq. (\ref{Omega}) corresponds to the
use of the macroscopic description of the electromagnetic wave.
The analysis of the microscopic structure of the electron-positron
pair production in plasmas is beyond the scope of the present
publication. However, since similar technique of the Lorentz
transformation into the reference frame, moving with the wave
group velocity, can be used to simplify the analysis of
electron-positron pair production inside the hollow waveguide and
in the focus region \cite{focus}, we think  that the main results
obtained above in the longwave approximation ($\left(2\pi
c/\omega\right)^3 n \gg 1$) will not change substantially. On the
other hand the elaboration of the microscopic description will
particularly determine the applicability of the approximation
used.

Using the properties of the dispersion equation for the
electromagnetic wave in plasma, we carried out the calculation of the
probability in the reference frame moving with the group velocity of the
wave. The great simplification of the problem has been achieved owing to the
fact that in this reference frame the magnetic field of the wave vanishes
and the problem has been reduced to the problem of the pair creation in
rotating electric field.

In the present paper we used "the imaginary time method" to
calculate the probability of pair creation by the circularly
polarized electric field. There arises a problem of validity of
the quasi-classical tunneling time and  "the imaginary time
method". The calculations of the pair production by the
time-varying electric field carried out within the frameworks of
"the imaginary time method" \cite{3} and Dirac theory \cite{NN}
agree to each other up to the preexponential factor accuracy. It
is also well known that "the imaginary time" technique has been
widely used in the problem of the atom ionization by strong
electromagnetic wave. Theoretical and experimental results
relevant to the ionization problem can be found in Refs.
\cite{delone,delone1} and in cited literature therein. See also
Ref. \cite{popov_ufn}.

We noticed that there is a drastic difference between the one
dimensional and the two dimensional cases of particle motion below
the energy barrier, i. e. in the one dimensional case the particle
emerges from under the barrier with zero momentum. In two
dimensional case the particle is produced with nonzero momentum
perpendicular to the instantaneous direction of the electric
field, which was first pointed out in Ref. \cite{PopovMarinov}.

In the moving reference frame we calculated the probability of the
pair creation. We found that one pair is produced in
$(10^{-6}\mbox{cm})^3$ volume in 500 fms when the field amplitude
reaches a value about 0.08 of the critical QED field for lasers
with the wavelength $\lambda=10^{-4}\div 10^{-8}$ cm. This value
of the field amplitude in plasma in the moving frame corresponds
to initial field amplitude in vacuum $E_0=E \omega/\Omega$. We
also studied the dependence of pair creation probability on the
wavelength of the electromagnetic wave.

We notice that the electron-positron pairs created via the
Schwinger mechanism can be distinguished from the pairs created in
the trident reaction during interaction of the plasma electrons
with nuclei as well as from the pairs produced by bremsstrahlung
photons. The electron-positron pairs created via the Schwinger
mechanism in the laboratory reference frame have the longitudinal
momentum larger than the pairs appeared due to the trident
reaction and bremsstrahlung induced pair production.

\section*{Acknowledgements}

The author would like to acknowledge many fruitful discussions
with V. S. Popov and to thank V. D. Mur, L. B. Okun, N. B.
Narozhny and M. I. Vysotsky for valuable remarks. This work was
accomplished in the framework of the Federal Program of the
Russian Ministry of Industry, Science and Technology N
40.052.1.1.1112. It was also partially supported by RFBR (grant N
00-15-96562).

\onecolumngrid{\newpage}

\section*{Figure Captions}

\noindent {\bf Fig.1} a) The dependence of the initial electron
(positron) momentum normalized on the amplitude of vector
potential s on the parameter $\gamma$; b) The dependence of the
electron sub-barrier motion time $\tau _{0}$ on the parameter
$\gamma$.

\bigskip

\noindent {\bf Fig.2} $g(\gamma )$ versus $\gamma$. The upper
curve corresponds to the circularly polarized electromagnetic
wave, which coincides with the result of Ref. \cite{PopovMarinov},
next one corresponds to the linearly polarized electromagnetic
wave $\sim\cos \omega t$, the last one also to the linearly
polarized wave, but with $1/\cosh \omega t$ time dependence.

\bigskip

\noindent {\bf Fig.3} a) The dependence of $g$ on $\varepsilon$
for different values of $\lambda$: $10^{-10}$ cm, $10^{-9}$ cm,
$5\times 10^{-9}$ cm, $10^{-8}$ cm, $10^{-7}$ cm, and $10^{-6}$
cm, from bottom to top; b) The dependence of $\log_{10}W$ in the
volume $(10^{-6}\mbox{cm})^3$ and in 500 fms on $\varepsilon$ for
different values of $\lambda$: $10^{-9}$ cm, $5\times 10^{-9}$ cm,
$10^{-8}$ cm, and $10^{-7}$ cm from top to bottom; c) The
dependence of $g$ on $\lambda$ for $\varepsilon=0.08$; d) The
dependence of $\log_{10}W$ in the volume $(10^{-6}\mbox{cm})^3$
and in 500 fms on $\lambda$ for $\varepsilon=0.08$.

\bigskip

\noindent {\bf Fig.4} a) The energy and the momentum of produced
pairs in the laboratory frame, averaged over time, versus
parameter $\gamma$ for $\omega/\omega_p=5$; b) The momenta of
"Schwinger" pairs (upper curve) and "trident" pairs in the
laboratory frame, averaged over time, versus parameter $\gamma$.

\onecolumngrid{\newpage}

\begin{figure}[tbp]
\begin{tabular}{ccc}
\epsfxsize6cm\epsffile{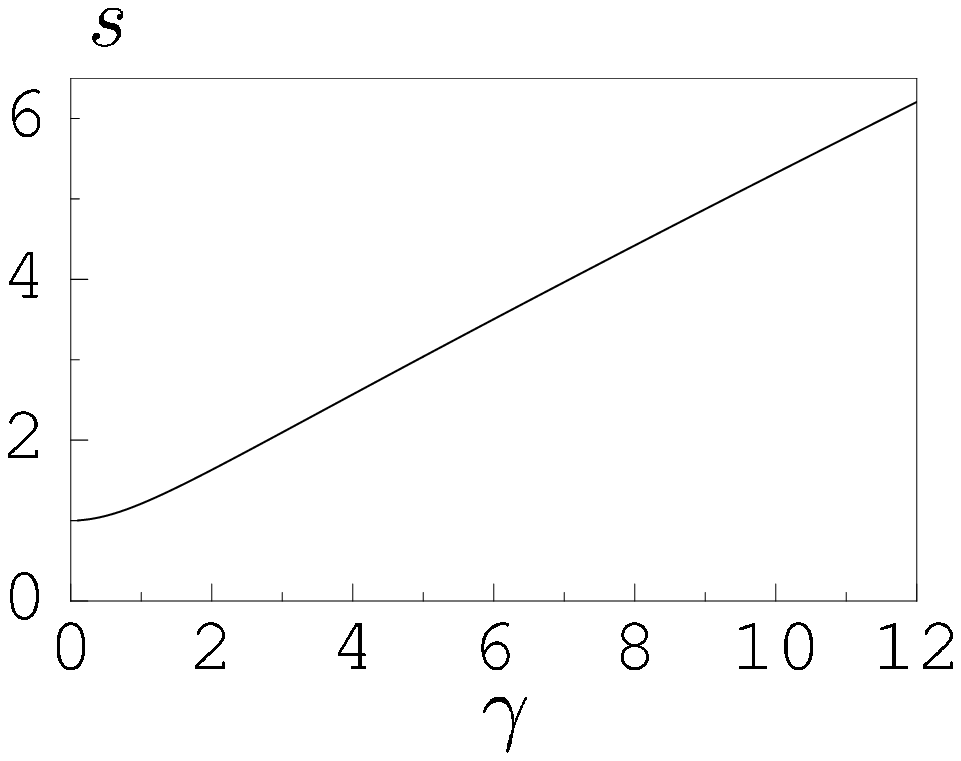} & \epsfxsize6cm\epsffile{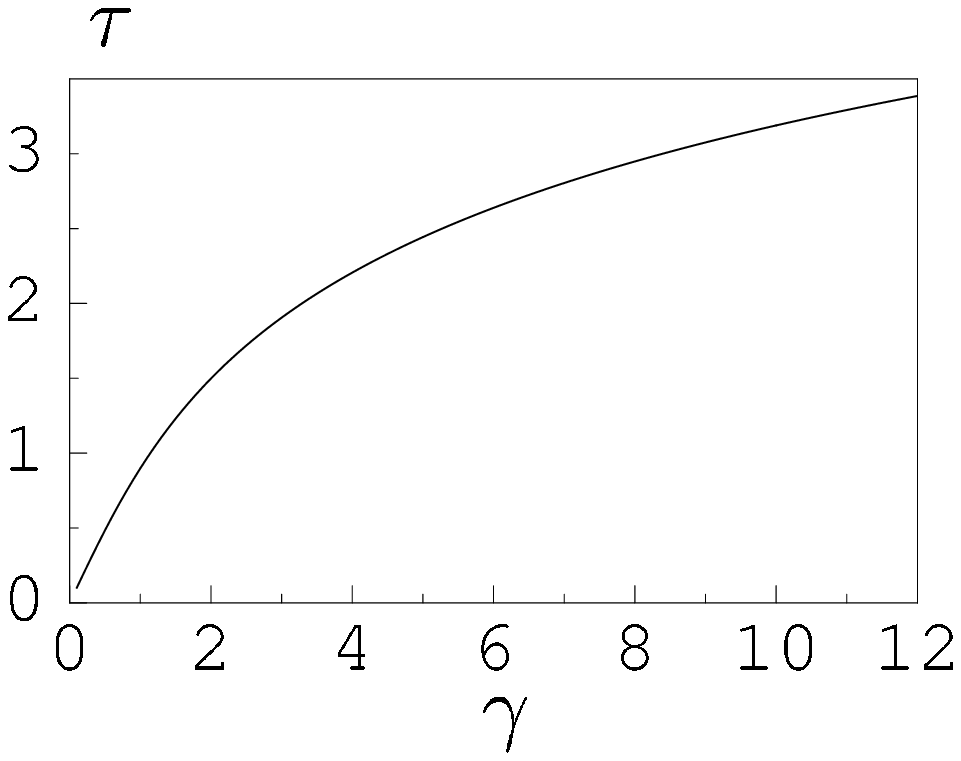} &
\\ a) & b) &
\end{tabular}
\caption{}
\end{figure}

\begin{figure}[tbp]
\begin{tabular}{ccc}
& \epsfxsize6cm\epsffile{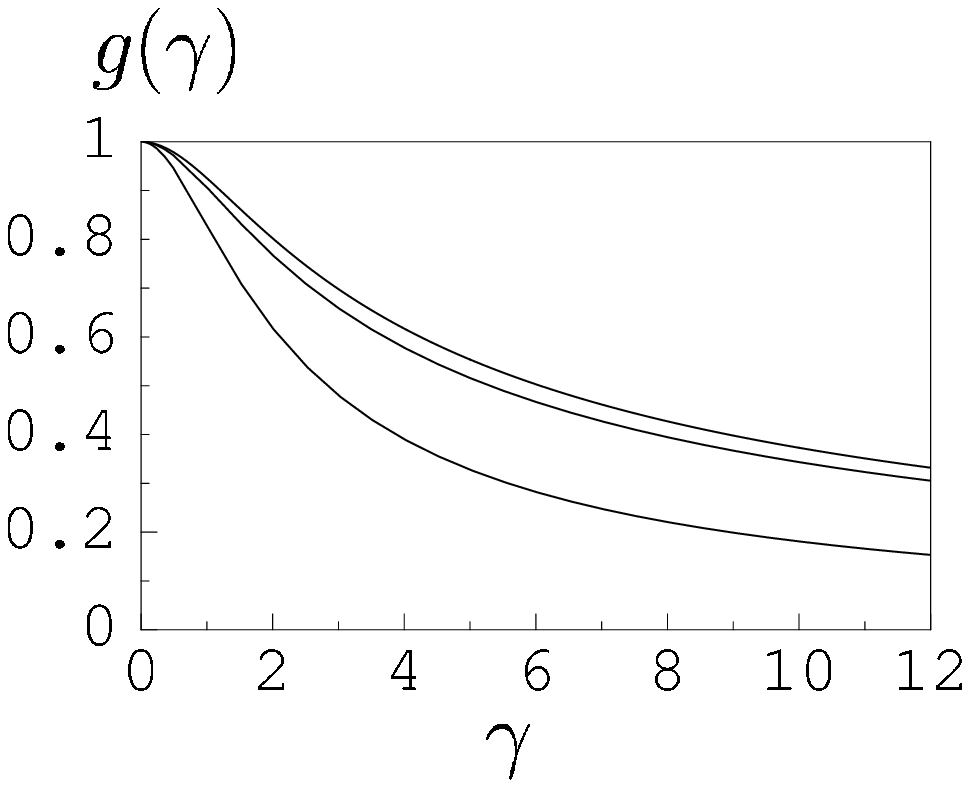} &
\end{tabular}
\caption{}
\end{figure}

\begin{figure}[tbp]
\begin{tabular}{ccc}
\epsfxsize6cm\epsffile{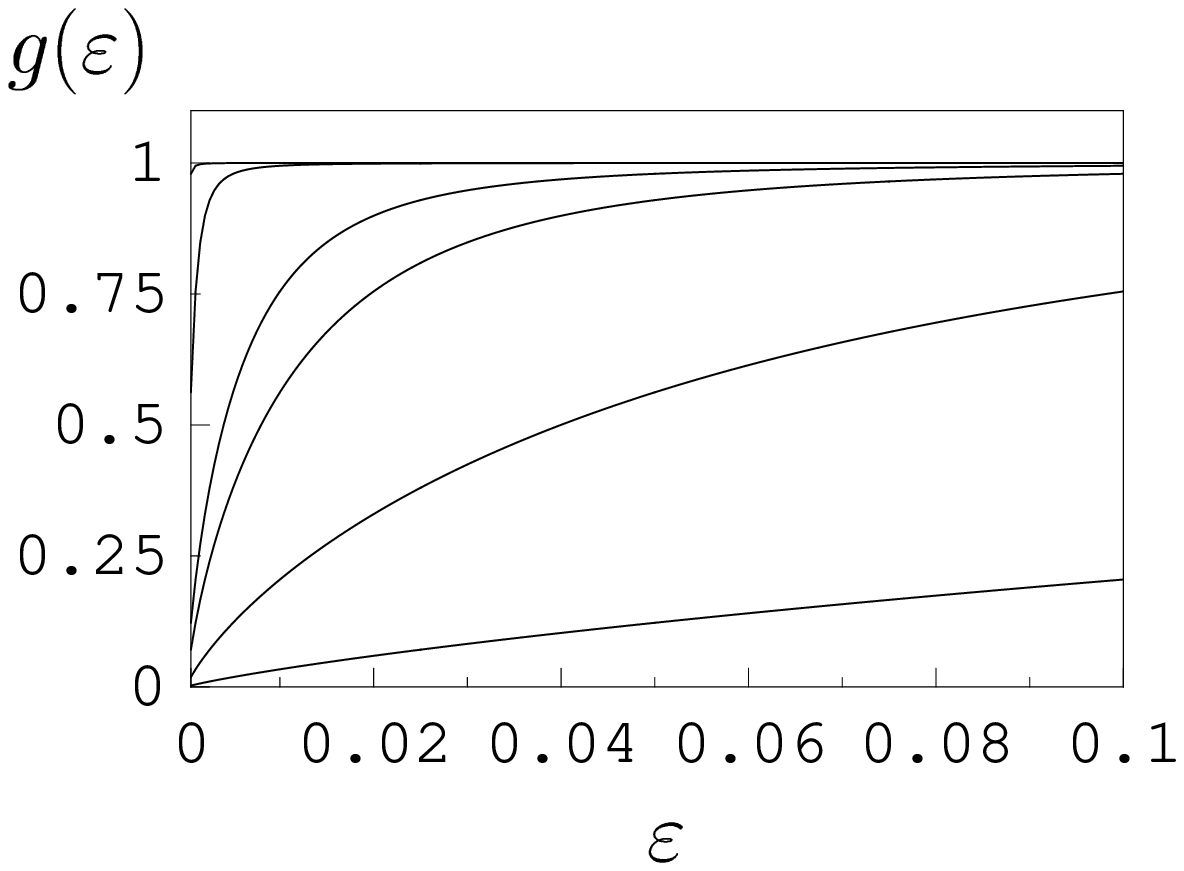} & \epsfxsize6cm\epsffile{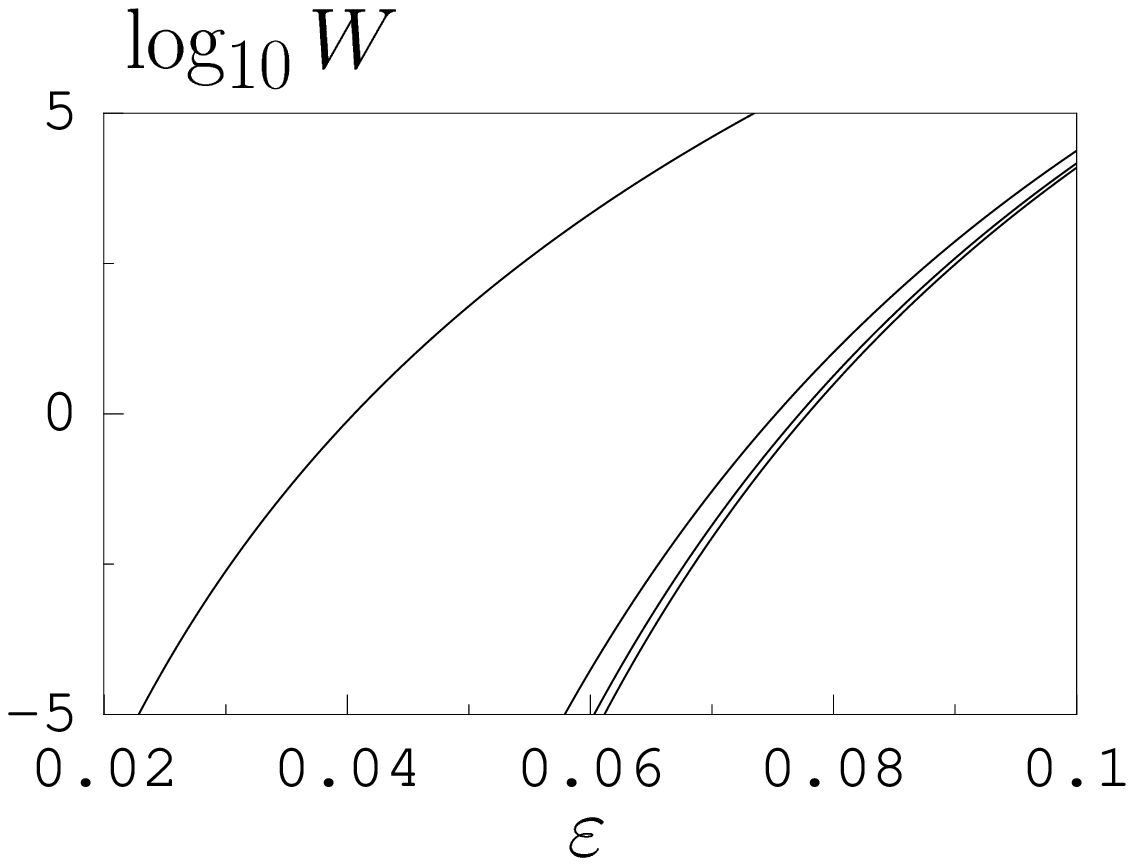} &
\\ a) & b) &\\
\epsfxsize6cm\epsffile{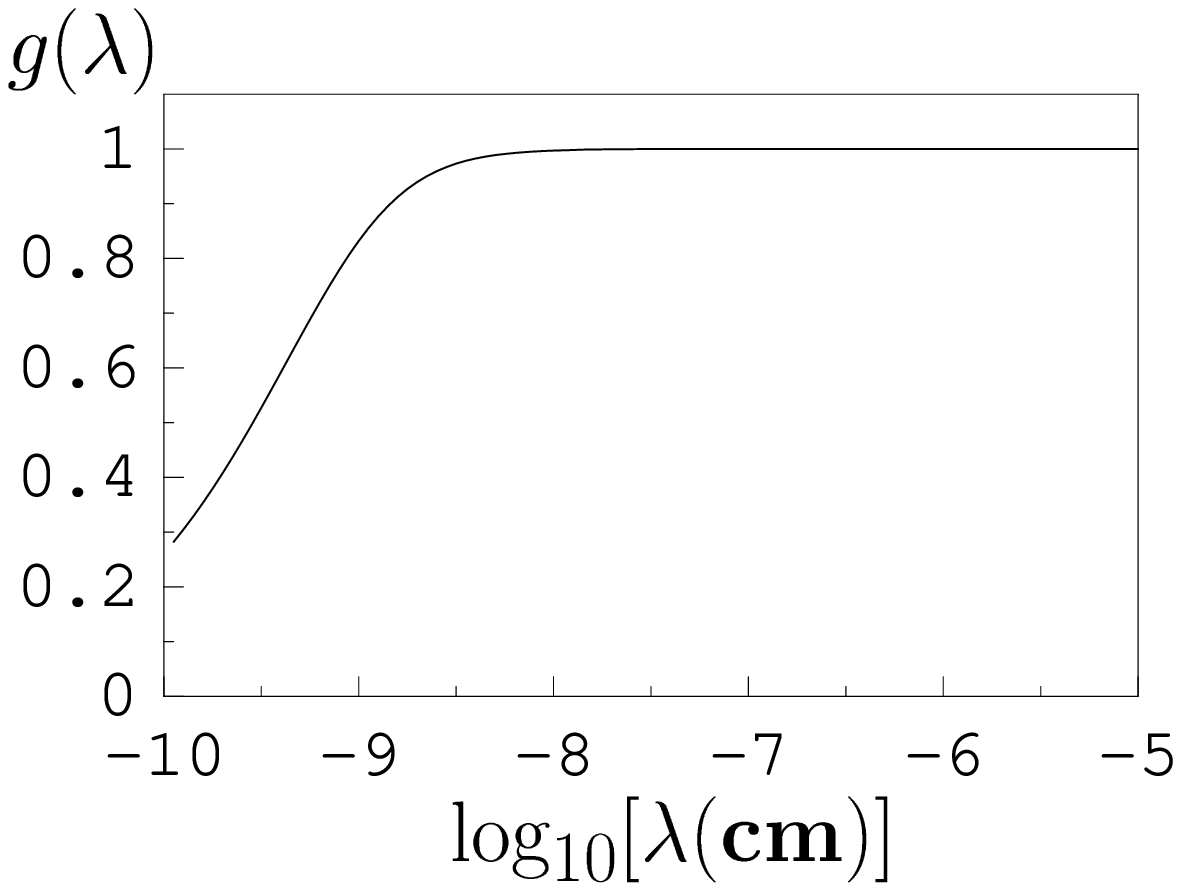} & \epsfxsize6cm\epsffile{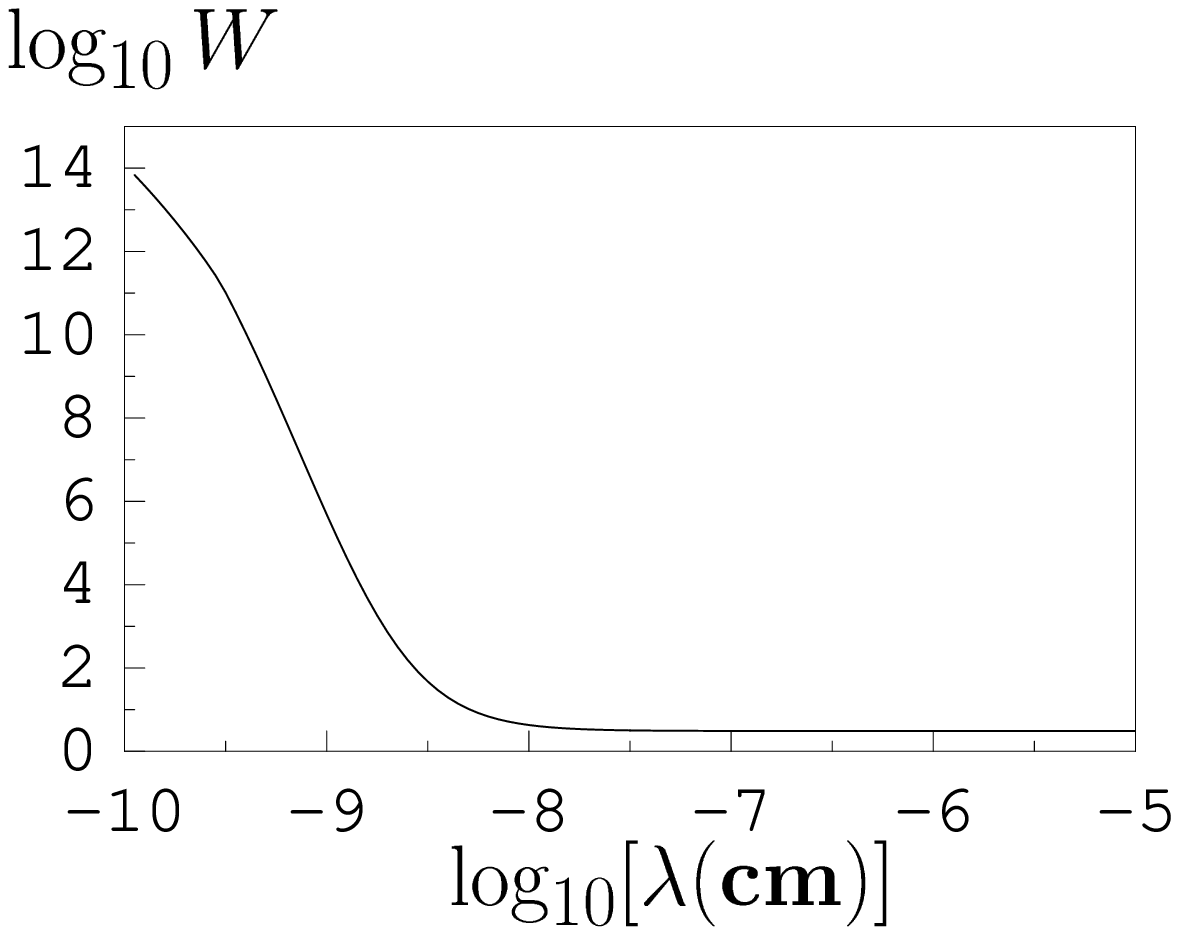} &
\\ c) & d) &
\end{tabular}
\caption{}
\end{figure}

\begin{figure}[tbp]
\begin{tabular}{ccc}
\epsfxsize6cm\epsffile{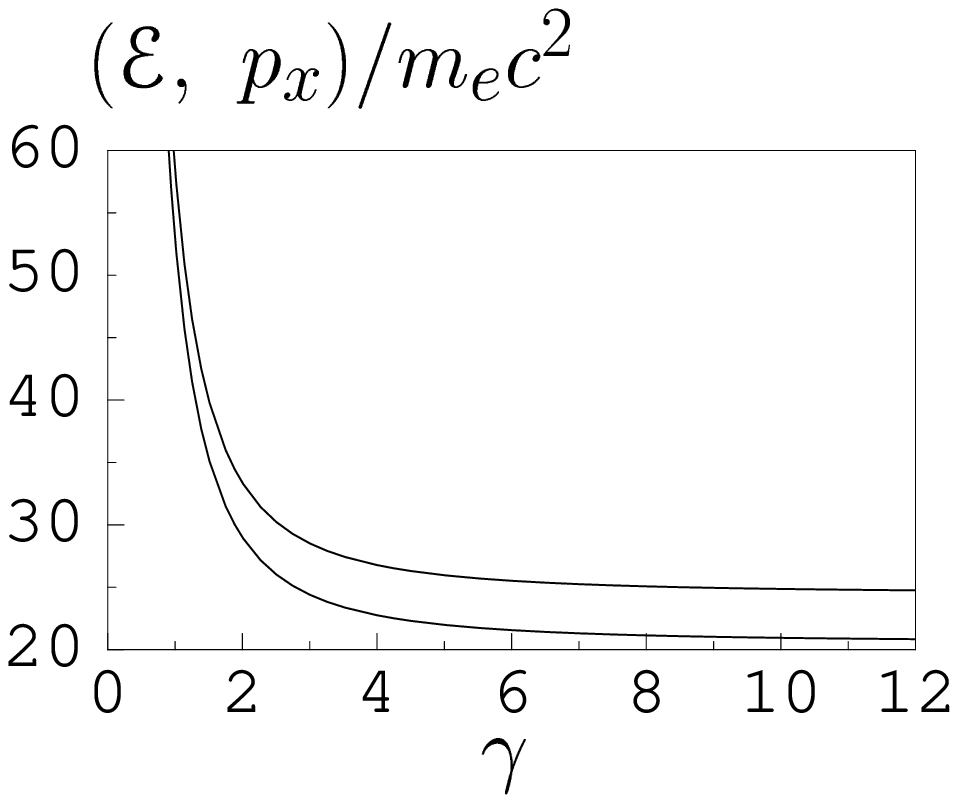} & \epsfxsize6cm\epsffile{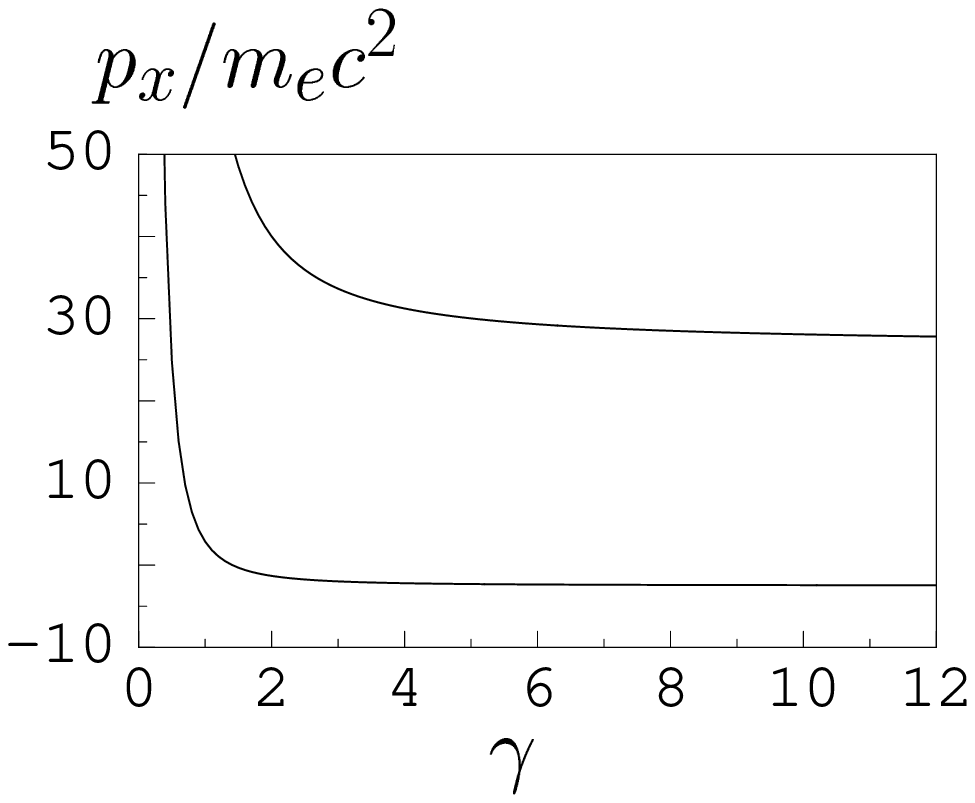} &
\\ a) & b) &
\end{tabular}
\caption{}
\end{figure}


\begin{thebibliography}{99}
\bibitem{Schwinger}  J. Schwinger, Phys. Rev. {\bf 82}, 664 (1951).

\bibitem{vacuum}  W. Dittrich, H. Gies, {\it Probing the quantum vacuum:
pertubative effective action aproach in quantum}. (Springer, New York:
2000).

\bibitem{el-pos}  W. Heisenberg and H. Z. Euler, Z. Phys. {\bf 98}, 714
(1936).

\bibitem{LandauLifshits-TF}  L. D. Landau and E. M. Lifshitz, {\it The
Classical Theory of Fields} (Pergamon Press, Oxford, 1980).

\bibitem{B-I}  E. Brezin and C. Itzykson, Phys. Rev. D {\bf 2}, 1191
(1970).

\bibitem{3}  V. S. Popov, JETP Lett. {\bf 13}, 185 (1971); Sov. Phys.
JETP {\bf 34}, 709 (1972).

\bibitem{4}  V. S. Popov, JETP Lett. {\bf 18}, 255 (1973); Sov. J. Nucl.
Phys. {\bf 19}, 584 (1974).

\bibitem{PopovMarinov}  M. S. Marinov and V. S. Popov, Sov. J. Nucl. Phys.
{\bf 16}, 449 (1973).

\bibitem{NN}  N. B. Narozhny and V. M. Frolov, Sov. Phys. JETP {\bf 38}%
, 427 ( 1974).

\bibitem{7}  V. M. Mostepanenko and A. I. Nikishov, Sov. J. Nucl. Phys.
{\bf 19}, 451 (1974).

\bibitem{8}  M. S. Marinov and V. S. Popov, Fortschr. Phys. {\bf 25}, 373
(1977).

\bibitem{9}  A. A. Grib, S. G. Mamaev, and V. M. Mostepanenko, {\it %
Vacuum Quantum effects in strong fields} (Energoatomizdat, Moscow, 1988).

\bibitem{Ringwald}  A. Ringwald,
Phys. Lett. B {\bf 510}, 107 (2001).

\bibitem{25}  F. V. Bunkin and I. I. Tugov, Dokl. Akad. Nauk SSSR {\bf 187%
}, 541 (1969).

\bibitem{26}  G. J. Troup and H. S. Perlman, Phys. Rev. D {\bf 6}, 2299
(1972).

\bibitem{27}  G. A. Mourou, C. P. J. Barty, and M. D. Perry, Phys. Today
{\bf 51}(1), 22 (1998).

\bibitem{SLAC99}  C. Bula, et al., Phys. Rev. Lett. {\bf 76}, 3116 (1996).

\bibitem{29} I. Flegel and J. Rossbach, CERN Courier {\bf 40}(6), 26 (2000); CERN Courier {\bf 41%
}(5), 26 (2001)

\bibitem{FEL}  R. Alkofer, M. B. Hecht, C. D. Roberts, S. M. Schmidt, and D. V. Vinnik, Phys. Rev Lett. {\bf 87}, 193902
(2001).

\bibitem{Tajima}  T. Tajima, Plasma Phys. Rep. {\bf 29}, 207 (2003).

\bibitem{sbul} S. V. Bulanov, T. Zh. Esirkepov, and T. Tajima, Phys. Rev. Lett. {\bf 91}, 085001 (2003).

\bibitem{Avetissian} H. K. Avetissian, A. K. Avetissian, G. F. Mkrtchian, and Kh. V. Sedrakian, Phys. Rev. E {\bf 66}, 016502 (2002).

\bibitem{PTP} A. M. Perelomov, V. S. Popov, and M. V. Terent'ev, Sov. Phys. JETP {\bf 24}, 207 (1968).

\bibitem{Popov}  V. S. Popov, JETP {\bf 94}, 1057 (2002).

\bibitem{trident}  J. W. Shearer, J. Garrison, J. Wong, and J. E. Swain, Phys. Rev. A 8, 1582 (1973).

\bibitem{AP}  A.~I.~Akhiezer, R.~V.~Polovin, Sov Phys. JETP {\bf 30}, 915
(1956).

\bibitem{Av}  H. K. Avetissian, A. K. Avetissian, A. Kh. Bagdasarian, and Kh. V. Sedrakian, Phys. Rev. D {\bf 54}, 5509
(1996).

\bibitem{prob1} S. P. Goreslavskii and S. V. Popruzhenko, JETP {\bf 83}, 661 (1996).

\bibitem{prob2} V. D. Mur, S. V. Popruzhenko, and V. S. Popov,
JETP {\bf 92}, 789 (2001).

\bibitem{focus} S. S. Bulanov, V. D. Mur, N. B. Narozhny, and V. S. Popov, in preparation.

\bibitem{keldysh} L. V. Keldysh, Sov. Phys. JETP {\bf 20}, 1307 (1964).

\bibitem{QED}  V. B. Berestetsky, E. M. Lifshitz, L. P. Pitaevsky, {\it Quantum Electrodynamics} (Pergamon Press, Oxford, 1982).

\bibitem{delone} N. B. Delone and V. P. Krainov, {\it Multiphoton Processes in Atoms} (Springer-Verlag, Berlin,
1994).

\bibitem{delone1} N. B. Delone and V. P. Krainov, Usp. Fiz. Nauk {\bf 168}, 531
(1998).

\bibitem{popov_ufn} V. S. Popov, Usp. Fiz. Nauk {\bf 19}, 819 (1999).

\end{thebibliography}
\end{document}